# 2024 'Key Reflections' on the 1824 Sadi Carnot's *'Réflexions'* and 200 Year Legacy

Milivoje M. Kostic

*Professor Emeritus* of Mechanical Engineering, *Northern Illinois University, USA*

Email: *kostic@niu.edu* * Web: kostic.niu.edu/ *&* http://Carnot.MKostic.com

**Abstract**

This author is not a philosopher nor historian of science, but an engineering thermodynamicist. In that regard and in addition to various philosophical *"why & how"* treatises and existing historical analyses, the physical and logical *"what it is" reflections, as sequential Key Points*, where *a key Sadi Carnot's reasoning infers the next one*, along with novel contributions and original generalizations, are presented. We need to keep in mind that in Sadi Carnot's time (early 1800s) the steam engines were inefficient (below 5%, so the heat in and out were comparable within experimental uncertainty, as if *caloric* were conserved), the conservation of *caloric* flourished (might be a fortunate misconception leading to the critical analogy with the waterwheel), and many critical thermal-concepts, including the conservation of energy (The *First Law*) were not even established. *Since Clausius and Kelvin earned to be "Fathers of thermodynamics," then Sadi Carnot was 'the ingenious' "Forefather of thermodynamics-to-become".*

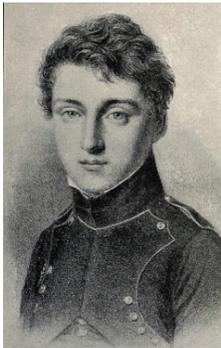

"*The motive power of heat is independent of the agents employed to realize it; its quantity is fired solely by the temperatures of the bodies between which is effected, finally, the transfer of the caloric.*" *– by Sadi Carnot, 1824* (English Translation by Robert H. Thurston [1]).

★★★

## 1. Introduction

In 2024 the thermodynamic community celebrated *Sadi Carnot's Legacy* and the *200$^{th}$ Anniversary* of his only and famous 1824 '*Réflexions*' publication, not appreciated at his time, and may be truly comprehended by a few, even nowadays [1-5]. Here we present this author's contribution with '2024 *Key*

*Reflections*' of his lifelong work and his plenary presentation at *Sadi Carnot's Legacy International Colloquium* at École Polytechnique in France [3, 6].

As Anthony Legget, a Nobel laureate, commented, "*Mathematical convenience versus physical insight […] that theorists are far too fond of fancy formalisms which are mathematically streamlined but whose connection with physics is at best at several removes […] heartfully agreed with Philippe Nozieres that 'only simple qualitative arguments can reveal the fundamental physics'* [*Sci. Bull.* 2018, *63*, 1019–1022]." In that regard, instead of extensive review of literature and 'extravagant' historical or philosophical formal methodology (as done by many); here, mostly thermodynamic, specific and simple reasoning, but with deep intuitive comprehension, were emphasized by a lifelong engineering thermodynamicist, for the first time in the literature as such.

It is well known that "*Thermodynamics is not easy to understand 'the first time around'*." Thermodynamics is confusing until it is comprehended, and this writing is no exception. Sometimes, interpretation by different experts may result in further confusion. Namely, not all scientists/physicists and engineers are thermodynamicists, and a "*thermodynamic historian*" is not a "thermodynamicist" but a "historian," often with elegant style and abstract methods, but sometimes without due comprehension of tacit thermodynamic fundamentals. We are often trapped in our own thoughts and words (especially in emerging new concepts and if nonnative) and the subtle holistic meanings are to be read "between the lines." Sometimes, highly accomplished scientists in their fields do not fully comprehend the essence of thermodynamics, especially if related to the *Second Law* and *entropy*.

The objective of this article is not to formally review the Carnot's '*Réflexions*' nor vast related literature by others (which is already, in part, done by this author [9, 10] and many others, for example [12-16]), but to present this author's long-contemplated reflections, with critical thermodynamic insights and logical reasonings, and to put "key thermodynamic concepts" in contemporary perspective. Namely, selective, physical and sequential '*Key Points*' as judged by this author, where ingenious Sadi Carnot's reasoning infers the next one, along with '*Miss Points*', presenting persistent post-misconceptions and fallacies by others (that needed to be underscored and '*put to rest*'), are presented for the first time as such. The emphasis is on engineering, phenomenological thermodynamics (on fundamental substance instead of formal methodology and style), and not on philosophical and historical review and 'extension' of Carnot work (already done by many). Therefore, only selected publications and related publications by this author are referenced.

Sadi Carnot, at age 28, published in 1824, now famous "*Réflexions sur la puissance motrice du feu"* (English translation, "*Reflections on the Motive Power of Fire* [1]"). His ingenious reasoning of reversible processes and cycles, and maximum "*heat-to-power efficiency*" laid foundations for the *Second Law of thermodynamics,* before *The First Law* of energy conservation was even formulated (in 1840s), and long before Thermodynamic concepts were established (in 1850s and later) [7, 8] and elsewhere. Sadi Carnot, who died in 1832 at age 36 from cholera epidemic, could not had been aware of immense implications of his ingenious reasoning at that time. No wonder that Sadi Carnot's masterpiece, regardless of flawed assumption of *conservation of caloric*, was not appreciated at his time, when his ingenious reasoning of ideal "*heat engine reversible cycles*" was not fully recognized, and may be truly comprehended by a few, even nowadays.

Before this author's "*2024 'Key Reflections' on the 1824 Sadi Carnot's 'Réflexions'* " are presented here, the brief introduction, based on his prior publications [9-11], is given next to revisit essential concepts. Note that Carnot's *Réflexions* focused on ideal, reversible cycles and maximum possible efficiency, hence, the reversible processes and cycles are assumed here unless specifically stated otherwise (as dissipative or irreversible). Furthermore, the concepts of *heat* and *work* were not well-formalized in the Carnot's time, but *caloric* and *motive-power* were used instead, the latter as work-rate or in short '*work*' is often used here (e.g., heat or work for duration of a cycle or for some time period, should be self-evident).

Sadi Carnot gave a full and accurate reasoning of heat engine cyclic-processes and their limitations of "*converting heat to* [work] *motive-power*" at the time when caloric theory was flourishing and almost two decades before equivalency between work and heat was experimentally established by Joule and others, in1840s, see elsewhere.

At that time, when the energy conservation law was not known and heat was considered as indestructible caloric, when heat engines were in initial stage of development with efficiency of less than 5%, the confusion and speculations flourished. Can efficiency be improved by different temperatures or pressures, a different working substance than water; or some different mode of operation than pistons and cylinders? With ingenious and far-reaching reasoning, Sadi Carnot answered all of those questions and logically reasoned (thus proved) that maximum, limiting efficiency of heat engine does not depend on medium used in the engine or its design, but only depends on (and increases with) the temperature difference between the heat source and cooling medium or heat sink (however not linearly), similarly to the water-wheel work-power dependence on the waterfall height difference at a given water flowrate, see *Equation* (1) and *Figure* 1 (explicit formulas were developed after Carnot followers' work [7, 8] and elsewhere; see also *Miss Point 1* and *Key NOVEL-Point 4*).

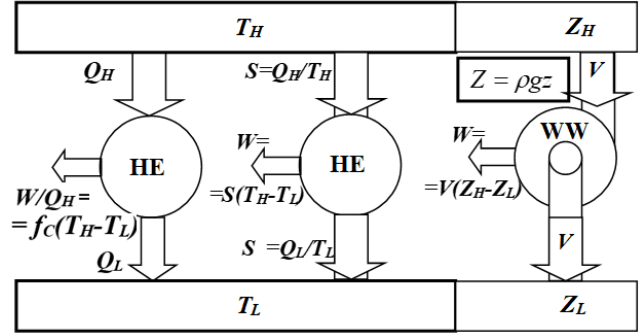

**Figure 1.** *Similarity between a heat engine* (HE) *and a waterwheel* (WW) [9]. At Carnot's time the '*caloric*' (heat) was considered to be conserved ($Q_H=Q_L=Q$ and due to very low heat engine efficiency at that time, about 3%, or $Q_L≈0.97Q_H≈Q_H$, so it was a reasonable misconception), leading to the *Waterwheel analogy*, with caloric-flow ($Q_{IN}=Q_H$) and high (*H*) and low (*L*) temperature-difference (although not linearly, see *Equation* (1), being the proof that Carnot was not and could not had been thinking even remotely of any *"other' caloric"* but heat, to imply "entropy-like quantity" as speculated by some, see *Miss Point* 4) corresponding to the water-flow and elevation-difference, respectively. It turned out, after discovery of entropy ($S=Q/T=$constant for reversible cycles [8]) that HE full similarity (with entropy and direct, linear temperature difference, *Figure* Center) with WW (*Figure* Right) was established.

The most importantly, Carnot introduced the reversible processes and cycles and, with ingenious reasoning of "*Contradiction impossibility*", see *Key NOVEL-Point 3,* proved that maximum heat engine efficiency is achieved by any reversible cycle, thus, all must have the same maximum-possible efficiency ([1, 9, 10], see also *Key NOVEL-Point 2*), i.e.:

*"The motive power of heat is independent of the agents employed to realize it; its quantity is fired solely by the temperatures of the bodies between which is effected, finally, the transfer of the caloric."* [1]. Namely,

$$W_C = W_{netOUT} = Q_{IN} \cdot f_C(T_H,T_L); \quad \eta_C = \left.\frac{W_{netOUT}}{Q_{IN}}\right|_{Max} = \underbrace{f_C(T_H,T_L)}_{Qualitative\ function}\bigg|_{Rev.} \quad (1)$$

Carnot cycle consists of four reversible processes, see *Figure* 2: isothermal heating and expansion at constant *high*-temperature $T_H$ (also referring to '*hot* reservoir' elsewhere; *process* 1-2); adiabatic expansion to achieve *low*-temperature $T_L$ (also $T_L \equiv T_C$ referring to '*cold* reservoir' elsewhere; *process* 2-3); isothermal cooling and compression at constant low-temperature $T_L$ (*process* 3-4); and adiabatic compression to achieve high-temperature $T_H$ and complete the cycle (*process* 4-1).

All processes are reversible; thus, the cycle could be reversed, without additional external intervention, along the same path and with the same quantities of all the heats and works in opposite directions (*in-to-out* and vice versa), see *Equation* (2) and *Figure* 3, i.e.:

$$\{Q_H, Q_L, W_C\} \underset{\text{IF REVERESED}}{\Leftrightarrow} \{-Q_H, -Q_L, -W_C\} \qquad (2)$$

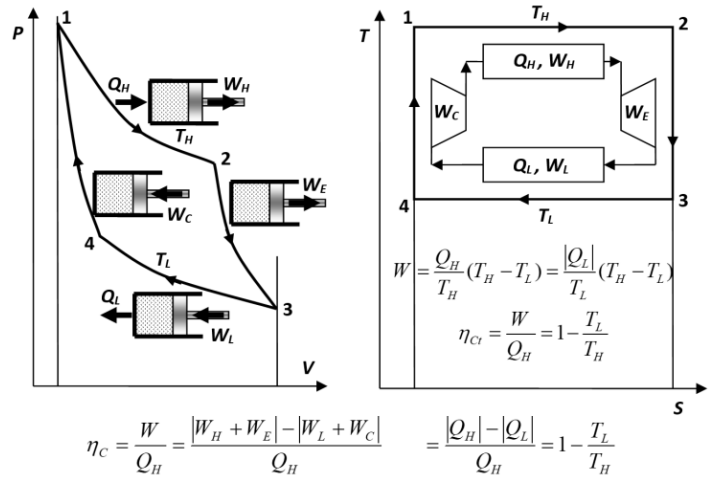

*Figure* 2. Heat-engine ideal-gas Carnot cycle [9]: note isothermal and adiabatic (mechanical) expansions (processes 1-2 and 2-3, respectively), and isothermal and adiabatic compressions (processes 3-4 and 4-1, respectively). The cycle net-work out is realized by isothermal processes ($W=W_H-W_L$) while adiabatic processes are needed to adjust temperatures with heat reservoirs to provide reversible, isothermal heat transfer, although their works cancel out ($W_E$-$W_C$ =0 [10]).

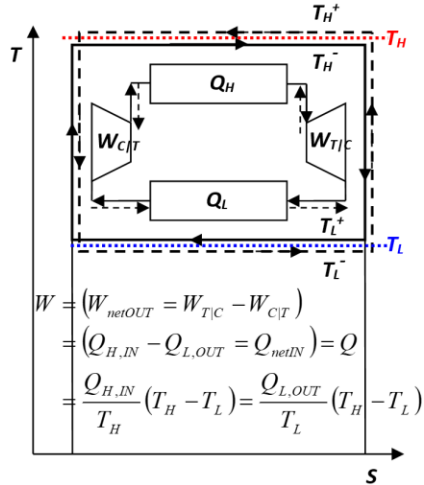

*Figure* **3.** Reversible steam-engine Carnot power-cycle (*solid lines*) and 'reversed', Carnot refrigeration-cycle (*dashed lines*, reversed directions) [9]. Note infinitesimal temperature differences above ($T^+$) or below ($T^-$) thermal reservoirs temperatures, $T_H$ (in *red*) and $T_L$ (in *blue*), to provide heating or cooling as needed.

The concept and consequences of a process and cycle reversibility are the most ingenious and far-reaching, see [5, 9, 10] (see also *Key NOVEL-Point 1*). Sadi Carnot's simple and logical reasoning that mechanical work is extracted in heat engine due to the heat passing from high to low temperature (see also *Key Point I*), led him to a very logical conclusion that any heat transfer from high to low temperature (like in a heat exchanger) without extracting possible work (like in a reversible heat engine) will be a waste of work potential — so he deduced that any heat transfer in ideal, perfect heat engine must be at infinitesimally small temperature difference (no loss of caloric-fall potential), achieved by mechanical compression or expansion of the working medium (required temperature adjustment without heat transfer), as Carnot ingeniously advised in full details [1] (see also *Key Point II*).

Then, Sadi Carnot expended his logical reasoning to conclude that all reversible (ideal) heat engines must have equal and maximum possible efficiency, otherwise if reversed, the impossible "*creation of conserved quantities*" would be achieved, see all details in [1, 9, 10] and elsewhere (see also *Key Point V* and *Key NOVEL-Point 2*). What 'a simple' and logical ingenious reasoning!

Carnot's reasoning proves that a reversible cycle cannot have smaller efficiency (power output relative to heat input) than any other cycle, thus all reversible cycles must have the same maximum possible power-efficiency for the given temperatures of the two thermal reservoirs, independently from anything else, including the nature of heat-engine design and its agent undergoing the cyclic process (see all relevant specifics in [1, 9, 10] and elsewhere; see also *Key Point V* and *Key NOVEL-Point 2*).

Since the irreversible cycles could not be reversed, they may (and do) have lower than maximum reversible efficiency up to zero (no net-work produced, if all work potential is dissipated to heat) or even negative (external work input required to run such a "parasite" engine which will dissipate such work, in addition to original work potential, into heat), i.e.:

$$\eta_{Irr} < \underbrace{\eta_{Rev} = \eta_{max} = f_C(T_H, T_L)}_{\text{Reversible}} \tag{3}$$

Carnot did not provide quantitative, but qualitative correlation for the ideal heat engine power-efficiency, and accurately specified all conditions that must be satisfied to achieve reversibility and the maximum efficiency: the need for "*re-establishing temperature equilibrium for caloric transfer,*" i.e. reversible processes, where the reversible heat transfer has to be achieved at negligibly small (in limit zero) temperature difference at both temperature levels, at $T_H$, high temperature for heat source (reversible heating), and at $T_L$, low temperature for heat sink (reversible cooling of heat-engine medium), see *Figures* 2 and 3; otherwise the work potential during heat transfer due to temperature difference will be irreversibly lost (the main Carnot's *cause-and-effect* reasoning), see also *Key Point II*.

Sadi Carnot reasoned that mechanical expansion and compression are needed to decrease and increase the temperature of the engine medium to match the low- and high-temperature of the thermal reservoirs, respectively, and thus provide for the reversible heat transfer [1].

Carnot then reasoned that in limiting cases, such as an ideal cycle, it could be reversed using the prior obtained work, to transfer back the caloric (heat) from low- to high-temperature thermal reservoirs, thus laying foundations for the refrigeration cycles (cooling and refrigeration/heat-pumps) as 'reversed' heat engine cycles, see *Figure* 3 and *Equation* (2).

Sadi Carnot's reasoning of "*heat engine reversible cycles and their maximum efficiency*" is in many ways in par with Einstein's *Relativity theory* in modern times, see *Equation* (4). It may be among the most important correlations in natural sciences that led to the discovery of *entropy* and the *Second Law* of thermodynamics, among others. This claim was stated by this author, 'symbolically' expressed by *Equation* (4), and named as '*Carnot* (ratio) *Equality*' in prior publications [9, 11] to emphasize invaluable but not-well recognized importance of *Q/T=constant* correlation, renamed here as '*Carnot-Clausius Equality*', as the precursor to and to resemble '*Clausius Equality*' cycle-integral, both formalized by Clausius [8], based on Carnot's Réflexions [1].

$$\left\{\underbrace{\overbrace{\frac{Q(T)}{Q(T_0)} = \frac{f(T)}{f(T_0)}\Big|_{f(T)=T} = \frac{T}{T_0} = \frac{Q}{Q_0}}^{\text{by Carnot's followers [Clausius (1854)]}}}_{\textbf{\textit{Carnot Equality (CtEq)}}} \text{ or } \frac{Q}{T} = \frac{Q_0}{T_0} \text{ i.e., } \frac{Q}{T} = constant \right\} \underbrace{\phantom{<\,?\,>}}_{<\,?\,>}^{\substack{\text{Essentially}\\ \text{Important as}}} \underbrace{\{mc^2\}}_{\substack{\text{Einstein}\\(1905)}}$$

(4)

The '*Key Reflections*' presented next are founded on the Sadi Carnot's '*Réflexions*' (English translation by Robert H. Thurston [1] and elsewhere) and on the developments of thermodynamics by the pioneers [7, 8] and others, emphasizing this author's views as a lifelong engineering thermodynamicist [5, 6, 9-11], as a complement to the existing science historians and science philosophers' analyses, e.g., [12-16].

Therefore, a key thermodynamic logic is used to recognize and infer the most probable sequential developments of Sadi Carnot's ingenious discoveries, as well as to reflect on the related analyses and misconceptions by others, considering the current state of knowledge; since now, we have the advantage to look at the historical developments more comprehensively and objectively than the pioneers. The sequential '*Key Points*', where key reasoning by Sadi Carnot infers the next one, along with '*Miss Points*' (persistent post-misconceptions and fallacies by others), including novel contributions and original generalizations by this author, as '*Key NOVEL-Points*' with '*Key Takeaways*', are presented next.

## 2. *Key Points*: Most probable sequential developments of Sadi Carnot's ingenious discoveries

  I. *The source of the heat engine "motive power" is "caloric fall" ("temperature fall" or temperature difference of the caloric)*
 II. *The "temperature fall," as source of engine motive power, should "not be wasted," but minimized in any "workless heat-transfer process"*
III. *All motive frictions and other dissipative processes should be minimized in order to maximize engine power and efficiency*
 IV. *Reversible Cycles: Isothermal heat transfer and other frictionless processes make an engine process or cycle reversible*
  V. *Reversible cycles must all have equal and maximum efficiency*

**Key Point I:** *The source of the heat engine "motive power" is "caloric fall" ("temperature fall" or temperature difference of the caloric)*

Hot *caloric* (*heat* at high temperature) is the cause and source of the *motive-power* (produced work) by the steam engines (the heat engines in general). Since at the time, the *caloric* was believed to be conserved (might be a fortunate misconception leading to the critical analogy with the waterwheel, see *Figure* 1); then, Sadi Carnot inferred that its hotness is producing the motive power while being cooled: "*motive power due to the [temperature] 'fall of the caloric'.*"

> **KEYNOTE 1:** All else the same, if temperature of heat is higher than the surrounding's, the higher 'temperature fall' through an engine, the more motive power will be (the more power per unit of heat flow, the more efficient engine), analogous to the more power from more "elevation-fall" from higher elevation through a water-wheel per unit of water flow. In Sadi Carnot's words, *"The temperature of the fluid should be made as high as possible, in order to obtain a great fall of caloric, and consequently a large production of motive power* [1]*."*

**Key Point II**: *The "temperature fall," as source of engine motive power, should "not be wasted," but minimized in any "workless heat transfer process"*

This is the most critical and ingenious reasoning by Sadi Carnot, *"wherever there exists a difference of temperature, motive power can be produced."* [1], that led to inference of ideal reversible cycles, the most critical concept of Carnot's discovery. If the temperature difference is the cause and source of motive power, then, if it is "consumed" during the heat transfer without work extraction, then its work potential would be lost, so the temperature difference during heat transfer should be minimized, i.e., be infinitesimal; ideally the heat transfer should be isothermal: *"need for reestablishing temperature equilibrium for caloric transfer … in the bodies employed to realize the motive power of heat there should not occur any change of temperature which may not be due to a change of volume* [1]*."*

**Key Point III:** *All motive frictions and other dissipative processes should be minimized in order to maximize engine power and efficiency.* Mechanical processes should be ideally frictionless to avoid work waste, i.e., dissipation losses.

**Key Point IV:** *Reversible Cycles: Isothermal heat transfer and other frictionless processes make an engine process or cycle reversible*

This is a monumental and crucial "*reversibility concept*" with far-reaching consequences. Reversible processes and cycles could effortlessly (without additional external compensation) reverse back-and-forth in perpetuity (like *perpetual motion*), therefore without any degradation or loss, being perfect and with maximum possible 100% efficiency. They take place at infinitesimal potential difference in either direction (in limit equipotential process, no potential loss of quality), without any quantity nor quality degradation, including conservation of the motive power or work potential. More details at *Key NOVEL-Point 1* and elsewhere.

**Key Point V:** *Reversible cycles must all have equal and maximum efficiency.*

This "*key discovery*" (*Carnot Theorem*) was ingeniously inferred by Sadi Carnot by logical reasoning that otherwise would result in creation of [assumed] conserved caloric and/or perpetual motion: *"the maximum of motive power resulting from the employment of steam is also the maximum of motive power realizable by any means whatever* [1]*."* This powerful insight is the most important and ingenious reasoning of Sadi Carnot, and it has far-reaching consequence as demonstrated much later by Kelvin [7] and Clausius [8], and other Sadi Carnot's followers. It is further elucidated in *Key NOVEL-Point* 2, and in *Key NOVEL-Point* 3 it is further generalized as "*Reversible Contradiction impossibility,*" see also *Figure* 4.

The selected and persistent post-misconceptions and fallacies by others are presented as '*Miss Points*' next:

## 3. *Miss Points*: Persistent post-misconceptions and fallacies by others

1. The well-known Carnot efficiency formula, $\eta_{Carnot} = W_{Rev|Max}/Q_H = (1 - T_L/T_H)$, was not established by Sadi Carnot, but much later by Kelvin and Clausius
2. The cause and source for motive power is the temperature difference, in principle, but not linearly dependent as misstated by some
3. The heat transferred out of the Carnot cycle at lower temperature is "not a waste heat" as often stated, but it is "useful quantity", necessary for completion of the cycle
4. Sadi Carnot could not had been thinking of "any other caloric" but heat, to imply "entropy-like quantity" as speculated by some

**Miss Point 1:** *The well-known Carnot efficiency formula, $\eta_{Carnot} = W_{Rev|Max}/Q_H = (1 - T_L/T_H)$, was not established by Sadi Carnot, but much later by Kelvin and Clausius*

Sadi Carnot inferred in 1824 the maximum heat-engine power efficiency as implicit function of thermal source-and-sink reservoirs' *high*-and-*low*, $t_H$ and $t_L$, temperatures only [$\eta_{Rev|Max} = W_{Rev|Max}/Q_H = f_C(t_H, t_L)$]. Note that absolute temperature concept was not known at that time. However, the well-known *Carnot efficiency* formula, $\eta_{Carnot} = W_{Rev|Max}/Q_H = (1 - T_L/T_H)$, with absolute temperature, sometimes attributed as developed by Sadi Carnot, was actually developed much later in 1850s, first by Kelvin [7] using ideal gas and later by Clausius [8] in general, and named "*Carnot efficiency*."

Paradoxically, it is shown here that Carnot, Kelvin and Clausius concepts of maximum, reversible cycle efficiency is misplaced, since fundamentally the Carnot cycle efficiency is not the "reversible cycle efficiency" *per se*, but a power-per-heat '*coefficient of performance, COP<1*', that include both, the heat engine and the thermal reservoirs efficiencies, to be decoupled, see *Key NOVEL-Point* 4. It is fundamentally like its inverse, heat-to-power *COP>1* of heat pump. Essentially, it is a thermal energy-source '*work-potential efficiency*', see more details in *Key NOVEL-Point* 4 and *Figure* 6. After all, the Carnot cycle efficiency does not depend on the cycle in any way, but it depends on the thermal reservoirs' temperatures only.

**Miss Point 2:** *The cause and source for motive power is the temperature difference, in principle, but not linearly dependent as misattributed by some*

Carnot stated that temperature difference is, in principle, cause and source for motive power, but not directly, not linearly dependent as misquoted by some [$W_{Max} = Q_H \cdot f_C(t_H, t_L) \neq f(t_H - t_L)$, not function of $\Delta t = t_H - t_L$ only]. As stated by Sadi Carnot, *"In the fall of caloric the motive power undoubtedly increases* *with the difference of temperature between the warm and the cold bodies; but we do not know whether it is proportional to this difference. ... The fall of caloric produces more motive power at inferior than at superior temperatures"* [1].

**Miss Point 3:** *The heat transferred out of the Carnot cycle at lower temperature is "not a waste heat" as often stated, but it is "useful quantity", necessary for completion of the cycle*

The heat transferred out of the ideal Carnot cycle at lower temperature is *"not a waste"* as often stated but it is necessary for completion of the cycle (the entropy balance), and therefore necessary and *useful quantity*. As stated by Sadi Carnot, *"... without 'the cold' the heat would be useless* [1] (see also *Figure 7*)."  The only waste is additional heat generated by irreversible work dissipation accompanied by *entropy generation* in real cycles, that must also be taken out to complete the cycle.

**Miss Point 4:** *Sadi Carnot could not had been thinking of "any 'other' caloric" but heat, to imply "entropy-like quantity" as speculated by some*

We need to keep in mind that in Sadi Carnot's time (early 1800s) the steam engines were inefficient (below 5%, so the heat-in and heat-out were comparable within experimental uncertainty, as if *caloric* is conserved), the *conservation of caloric* flourished, and many critical thermal-concepts, including the conservation of energy (*The First Law*) were not even established. At that time the *entropy* concept was not known even remotely. Therefore, Sadi Carnot could not had been thinking of "any '*other*' caloric" but heat, to imply "entropy-like quantity" as speculated by some, see *Key NOVEL-Point 6* for more details.

Novel contributions with deeper physical insights and related generalization by this author are formalized in the following '*Key NOVEL-Points*':

## 4. *Key NOVEL-Points*: Novel contributions and original generalizations

1. "Reversible and Reverse" Processes and Cycles Dissected
2. Maximum Efficiency and "Reversible Equivalency" Scrutinized
3. Reversible Contradiction Impossibility ("Reductio ad absurdum")
4. Reversible Carnot Cycle Efficiency Is Misplaced — It is NOT the "Cycle efficiency" 'per se', but a "Thermal energy-source 'work-potential efficiency'"
5. The Carnot-Clausius [Ratio] Equality (CCE) and Clausius Equality (Cyclic integral) are special cases of relevant "Entropy boundary integral" for reversible stationary processes
6. The 'caloric' is transformable to work and cannot be 'extended and renamed as entropy' which is 'the final transformation'

**Key NOVEL-Point 1:** *"Reversible and Reverse" Processes and Cycles Dissected*

Ideal and perfect, *"Reversible processes"* take place at infinitesimal potential difference (temperature, pressure and similar) at any instant within and between a system and its boundary surroundings, but they may and do change in time (*process* is a change in time). Namely, the spatial gradients are virtually zero at any instant while time gradients and related fluxes may be arbitrary as driven by ideal boundary surroundings and facilitated by ideal arbitrary (or infinite) transport coefficients. Therefore, the potential qualities of flux quantities (heat and different kinds of works) are not degraded but equipotentially transferred and stored between the system and its boundary surroundings, and thereby 'truly' conserved in every way. However, in time, due to unavoidable irreversible dissipation of work to generated-heat, accompanied by generation of entropy, all real processes between interacting systems (including relevant surroundings) are asymptotically approaching common equilibrium with zero mutual work potential and maximum mutual entropy.

Namely, if an elastic, ideal gas or ideal spring is *reversibly* compressed, then the pressure may change in time but is equal everywhere at any instant across the system and the boundary surroundings (equipotential driving force at any instant). Similarly, if heat is *reversibly* transferred, the temperature may change in time, but it is equal everywhere within the system at any instant, and if it varies in time, it is driven by varying, but spatially equipotential surrounding temperature, so the energy potential quality is stored and conserved everywhere in every way (it may be reversed *back-and-forth in perpetuity* without additional external compensation).

Note that 'time and energy-rates' are irrelevant *per se* for the reversible analysis of energy balances and properties between initial and final states, being independent of process type and path of how the final state is achieved, either reversibly or irreversibly, the former being more simple and suitable for analysis than the latter.

A *Cycle* is a special case of *quasi-stationary process* when flow inlet and outlet quantities are the same (feed into each other) and close the cycle. Like a stationary process, a cycle does not accumulate flux quantities and may repeat and last in perpetuity (quasi-stationary). Note that all processes, particle-wise, are transient in time (in *Lagrangian* sense, from inlet to outlet), but for the steady-state or stationary processes (in *Eulerian* sense) the properties do not change in time (zero temporal gradients) at a fixed location, and for a cyclic process the flow inlet quantities are the same as the outlet's (since they feed into each other).

A "*Reverse*" concept is independent of and should not be confused with the *reversible* concept. If reversible, a *reverse-process* could be reversed using prior, related process work (with infinitesimal change of potential difference in opposite direction, and without additional external-work compensation), while to reverse an *irreversible* process it would require additional, external work compensation.

A "*Reverse process*" and/or "*Reverse cycle*" would take place if the driving (forced) potentials of a reversible process or a cycle are reversed (by infinitesimal change in opposite direction), then such a reversible process would be reversed with all quantities changing direction from input to output (and vice versa; e.g., a *refrigeration cycle* is a *"reverse"* of a *power cycle* or vice versa, see *Equation* 2 and *Figure* 3). For stationary processes, no temporal gradients, there is no accumulation of flux-quantities within a system, and for quasi-stationary cycles no accumulation of flux-quantities after completion of a cycle. The input and output quantities would be conserved and could be *reversed back-and-fort in perpetuity* like perpetual motion.

In reality, there is a need for at least infinitesimal temperature difference (and/or pressure and similar) to provide a *process 'sense of direction'*, and to resolve directional ambiguity by chance. Therefore, every process must be at least infinitesimally irreversible (infinitesimally imperfect), the *reversibility* being an asymptotic, limiting ideal concept. For this reason, even reversible equilibrium is unachievable, like absolute zero temperature or any other ideal concept, only approached asymptotically.

**Key NOVEL-Point 2:** *Maximum Efficiency and "Reversible Equivalency" Scrutinized*

Maximum efficiency of an energy process or cycle entails *maximum-possible work extraction* from a system while coming to equilibrium with a reference system, usually the surroundings; or *minimum-possible work expenditure* in a *reverse process* of formation of original system (from within the same reference state), see *Figure* 4 (Center). Since the reversible processes do not degrade any potential quality and could be reversed without external intervention, the two works must be the same, with opposite sign only, for all reversible processes, and they represent the maximum *work potential* (WP) for the given nonequilibrium conditions. Therefore, the reversible processes are perfect, and *equally and maximally 100% efficient*. They define the concept of *"Reversible Equivalency"* — the *'true quantity and quality equality'* of input and output, where relevant quantities and qualities are conserved in perpetuity. In real processes there will be some work dissipation losses (degradation of work with its dissipative conversion to heat), so that *less work* would be extracted than the maximum possible, and *more work* would be needed than the minimum required, thus reducing the maximum possible efficiency for real, *irreversible processes,* see *Equation* (3) and *Figure* 4 (Left).Since the reversible processes take place at virtual equipotential condition (virtually the same temperature, pressure, etc.; they are equipotential locally at any time, thus being reversible at any time), but they may vary in time with time-variable system and surrounding properties. Therefore, 'potential quality' of all relevant quantities would be equipotentially transferred and stored at any time, i.e., conserved without any degradation (without any dissipation) and could be effortlessly reversed back-and-forth (without additional external compensation).

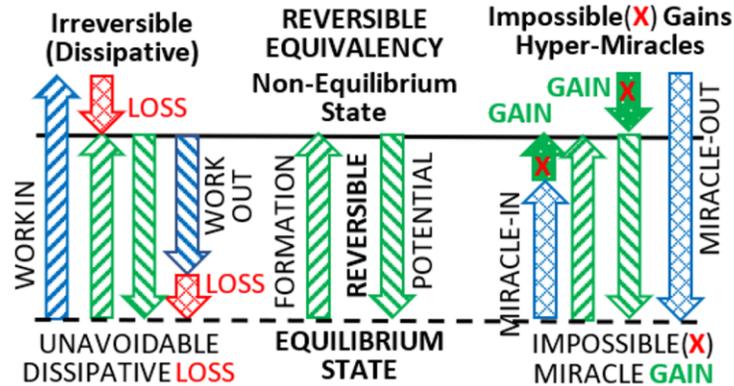

*Figure* **4.** *Reversible Equivalency and Contradiction Impossibility*: Ideal *Formation-work* of "non-equilibrium state" (minimum required work with perfect efficiency) is equal to its *Work-Potential,* WP (maximum available work with 100% perfect efficiency (No LOSS; *Center*), otherwise it will result in *Contradiction impossibility (Right)*. Formation of non-equilibrium state with less than its WP or getting more useful-work than WP would require a "*miracle work-GAIN*" without due work-source (violation of *Second Law*), being against natural-forcing and existence of equilibrium, thus impossible (*Right*). However, real formation *Work-In* is bigger and retrieved useful *Work-Out* is smaller than its stored WP (*Left*) due to unavoidable dissipation LOSS, resulting in smaller than 100% perfect efficiency (*Left*). Furthermore, any ideal, reversible 'underachievement' when reversed would become 'overachievement', would form WP with less work, resulting in impossible, miracle GAIN, as well as 'overachievement' (over WP; *Right*), i.e., all reversible processes must be equally perfect with 100% efficiency (*Center*).

With infinitesimal reversal of relevant potentials, all flux quantities will change directions while conserving the quantities and qualities. Therefore, the work extraction in a *reversible cycle* (i.e., Work-Potential, WP ) would be equal to the expenditure or Formation-Work in its '*reverse*' *reversible-cycle*, see *Equation* (2) and *Figures* 3 & 4. Furthermore, all reversible cycles must have equal and maximum 100% efficiency, otherwise any *'under-achieving'* reversible cycle (with lower work extraction than another [reference] cycle), when reversed, would consume less work than the reference cycle and thus be *'over-achieving'* with higher than reference 100% maximum efficiency, resulting in a *'Contradiction impossibility'*, see *Key NOVEL-Point* 3 next.

**Key NOVEL-Point 3:** *Reversible Contradiction Impossibility ("Reductio ad absurdum")*

As stated above, the reversible efficiency implies *maximum* work extraction and *minimum* work expenditure in the reverse process, thereby the two must be equivalent for given conditions, thus establishing the *Reversible Equivalency*, see also *Key NOVEL-Point 2*. Otherwise, any reversible cycle "*under-achievement*" (getting less than maximum possible) would become an *"over-achievement"* when such cycle is reversibly 'reversed' (accomplishing with less than minimum required), and such "*reverse over-achievement*" would be physically impossible, would be the "*Reversible equivalency* violation" and may violate the *conservation law*s, thus implying the *"Reversible Contradiction Impossibility"* of a well-known fact, see *Figure* 4-Right (impossible *'miracle* GAIN'*).

Namely, the *"Reversible Contradiction Impossibility"* (an *under-achieving* reversible process when reversed would become an *impossible over-achieving* reversible process) could result in numerous consequences: Namely, miraculous creation of 'perpetual motion' or creation of assumed 'conserved caloric' (regardless of Sadi Carnot's misconception of *caloric* conservation), and other impossible processes, like spontaneous heat transfer from lower to higher temperature, etc. The 'reversible contradiction impossibility' is so strong and universal a concept that any pertinent or quasi-relevant criteria, even if misunderstood, like conservation of caloric, will be sufficient to reason fundamental inferences [1].

Further consequences of the *"Reversible Contradiction Impossibility"* would be spontaneous generation of a conserved quantity, or generation of nonequilibrium work potential, or energy transfer from lower to higher potential, like spontaneous heat transfer from lower to higher temperature and generation of thermal nonequilibrium, i.e., impossible destruction of entropy, see more details in [9-11] and elsewhere. Or, in general, spontaneous creation of non-equilibrium from within an equilibrium being the physical contradiction of always observed "spontaneous process, forcing-direction from non-equilibrium towards mutual equilibrium," and never experienced otherwise. It will amount to the *"forced-directionality contradiction"* of the *irreversible process-directionality* from a higher to lower potential towards mutual equilibrium, as well as *impossibility to reverse dissipation*.

> **KEYNOTE 2:** It would be logically and otherwise impossible and absurd ("*Reductio ad absurdum"*) to have a spontaneous process "*the one way and/or the opposite way*" arbitrarily in opposite directions, as casual by chance (i.e., to have heat transfer "*from hot-to-cold* or *from-cold-to-hot"* or *"forcing in one direction and accelerating in opposite direction*," by chance). *It would be a violation of the Second law of thermodynamics (2LT).*

The reversible processes are equipotential and therefore do not degrade nonequilibrium, but store and/or convert one kind to another, like a reversible cycle converts '*heat at high temperature*' to '*work plus heat at lower temperature*', and in reverse in perpetuity, defined in [11] as "*Carnot-Clausius Heat-Work Equivalency, CCHWE"* ('potential-like' heat at high temperature equivalent and converts to 'kinetic-like' work plus heat at lower temperature, and vice versa, analogous to a reversible pendulum, converting potential to kinetic energy, and in reverse, in perpetuity), see *Figure* 5. The *CCHWE* is a fundamental and autonomous physical concept, independent of any process or device [11].

Therefore, all reversible processes are perfectly "equivalent in every way" and the most efficient, without any dissipative degradation.

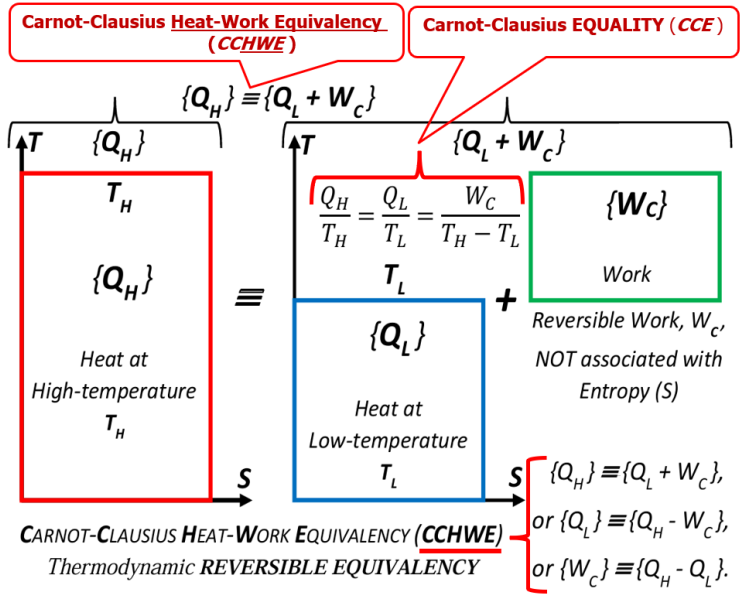

*Figure* **5.** The *heat* may in part be converted to *work* and vice versa (in "*thermal transformers*" i.e., heat engines and heat pumps), i.e., heat and work are reversibly-transformable (interchangeable as equivalent), generalized and named as "*Carnot-Clausius Heat-Work Equivalency (CCHWE),*" as "'*Heat at high temperature*' is (reversibly) equivalent, i.e., transformable, to '*work plus heat at low temperature*'". The *CCHWE* is a fundamental and autonomous physical concept, independent of any process or device [11]. The *Q/T=constant*, is invaluable but not-well recognized correlation, named here "*Carnot-Clausius Equality (CCE)"* as the precursor to and to resemble '*Clausius Equality*' cycle-integral, both formalized by Clausius [8], based on Carnot's Réflexions [1], see *Equation* 4.

.

**Key NOVEL-Point 4:** *Reversible Carnot Cycle Efficiency Is Misplaced – It is NOT the "Cycle efficiency" 'per se', but a "Thermal energy-source 'work-potential efficiency'"*

The *reversible processes and cycles, as a matter of concept, are 100% perfec*t without any degradation and must be equally and perfectly (maximally) efficient, *not over nor below* 100% efficient (would be the *Reversible Contradiction Impossibility*). Therefore, all reversible processes and cycles have 100% "*true quantity and quality*" efficiency: they extract 100% of "available *work potential"* as does any ideal waterwheel and any other reversible *energy-transformer* (e.g., engine or motor). The 100% perfect "*true reversible efficiency*" [11 *(CCHWE)*], see *Figure* 6-Right, should not be confused with "*maximum work-thermal efficiency*" of a thermal energy source, that represents the "*work potential of heat*" or *Exergy* of heat (or nonequilibrium thermal energy) of the relevant thermal reservoirs [$E_x=W_{Rev|Max} = Q(1-T_0/T_H)$, where $T_C=T_0$], see *Figure* 6-Left, as their property-like quantity, being independent of any heat engine or energy device

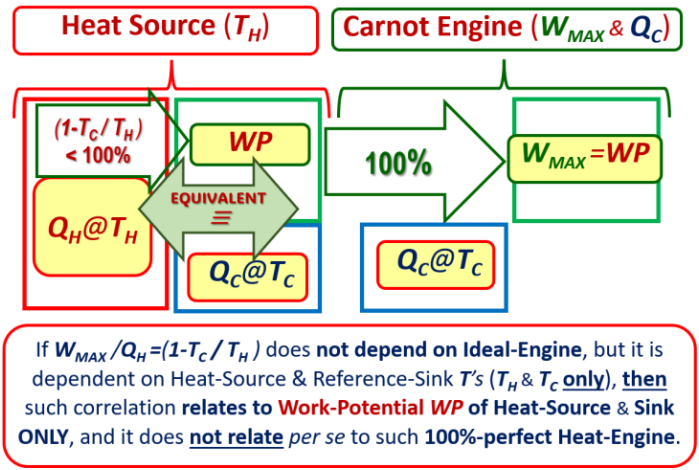

*Figure* **6.** *Carnot heat-engine efficiency is misplaced*: The *reversible processes and cycles (power and reverse), as a matter of concept, are 100% perfec*t without any degradation and must be equally and perfectly (maximally) efficient, not over nor below 100% ('*Reversible equivalency*', see *Figure* 4). Carnot defined engine efficiency, logically and "empirically," as "[work] motive-power *output per* heat *input*," thus implying both, engine and heat reservoirs combined system. Since ideal engines are maximally 100% efficient (*Figure* Right) and overall efficiency depends on the reservoirs' temperatures only, not on the engine in any way, then famous ($1-T_C/T_H$) Carnot efficiency refers to the former only (*Figure* Left). However, it would be hard "*to let go*" of the 200-yearlong "habit and addiction."

Sadi Carnot [1] and his followers, including Kelvin [7] and Clausius [8], ironically referred to the maximum heat-engine cycle efficiency (they "agonizingly" developed at the time when most thermal concepts were unknown), with the *absurd conclusion*, that "*it does not depend on the cycle design itself nor its operation mode* [in any way whatsoever]," hence not related to the cycle in any way, the proof that it is not the efficiency of ideal Carnot cycle *per se*. Therefore, their attribution is misplaced since the efficiency they developed *should had referred to* the "maximum motive power or *'work potential (WP)'* of the thermal reservoirs" since it depends on their temperatures only, and hence, being the logical proof of the claim presented here.

> **KEYNOTE 3**: Carnot heat-engine efficiency dependance on temperatures of the heat reservoirs only would be equally misplaced as if to attribute the maximum efficiency of an ideal waterwheel (water turbine), based on its motive-power per unit of input water-flow, and then it would also mistakenly depend on the water-reservoirs' elevations only. Therefore, all reversible devices are and have to be equally (not below nor above) the maximum possible 100% efficiency.

A motive power efficiency (i.e., a device's *work efficiency*) should be consistently based on the *work potential* of an energy source (not on a "convenient nor arbitrary input quantity," like heat input or water-flow input, etc.); and then, the 'true' *Carnot cycle efficiency* would be 100% as for all other efficiencies for ideal, reversible engines and motors.

Sadi Carnot defined engine cycle efficiency, logically and "empirically," as "[work] motive-power *output per* heat *input*," long before the concept of "*work potential*" of an energy-source and *energy conservation* were established. However, we now have the advantage of looking at the historical developments more comprehensively and objectively than the pioneers [5, 9-11] and to put in order the historical misconceptions.

An exact "*reverse*" of the reversible "*Power Carnot-cycle*" is the ideal "*Heat-pump cycle*" ("*Reverse Carnot-cycle*") whose efficiency or "performance" is defined 'in inverse' as "heat *output per* work *input*." It is always over 100% (as the "fundamental inverse" of the *Carnot cycle efficiency,* latter always smaller than 100%); and it is named as the *"Coefficient of Performance (COP)"* since 'such efficiency' over 100% would not be fundamentally (nor "politically") proper, but it should be fundamentally the same, i.e., improper if smaller than 100%.

> **KEYNOTE 4:** For the same fundamental reason, the efficiency of a perfect, *ideal* Carnot cycle (being below 100%) would also be logically inappropriate (as if there are "some *work losses*" in the ideal reversible cycles). For the same reason, as for the *Heat-pump cycle*, it should be called '*Carnot cycle COP*' but *not the 'Cycle efficiency'*. Fundamentally, all ideal, reversible cycles must be "*equally and maximally [100%] efficient*," as reasoned by Sadi Carnot [1].

Furthermore, it is fundamentally inappropriate, as often stated, to call the heat transferred out of the Carnot cycle at lower temperature, the "*waste heat or loss*", since it is the "*useful quantity,*" *necessary* for the completion of the perfect, ideal cycle, and together with the cycle work, they are reversibly-transformable (interchangeable as equal) and present the "*reversible equivalent*" to the heat-input at the high temperature, named by this author as *Carnot-Clausius Heat-Work Equivalency* (*CCHWE*) [11]), crucially fundamental and autonomous concept (independent of any process or device), see also *Figure* 5. The only "waste or loss" that lower efficiency below 100% would be any '*irreversible work dissipation*' (converted into generated-heat) and accompanied by *entropy generation*, that must be also taken out to complete a real cycle. A device's efficiency should not be higher than 100%, and only it could be lower due to irreversible, dissipative losses.

The "original," nowadays well-known *Carnot cycle efficiency* is misplaced and inappropriate, and it should be renamed for what it is: the *Work potential (WP*, o*r 'available energy') efficiency* of a heat-source-and-sink, or *Exergy efficiency* of a thermal-energy source with respect to the heat-sink reference. We now know that "true" Carnot efficiency, the *Second-Law* or *Exergy efficiency* is 100%. It is a goal here to clarify and resolve what is fundamentally misplaced. However, it would be hard "*to let go*" of the 200 yearlong "habit and addiction."

**Key NOVEL-Point 5:** *The Carnot-Clausius [Ratio] Equality (CCE) and Clausius Equality (Cyclic integral) are special cases of related "Entropy boundary integral" for reversible stationary processes*

The balance equations (used for definition of a new property, the *entropy*) were first developed by Clausius, based on Carnot's discovery of "maximum efficiency and equality for all reversible cycles," named here, *Carnot-Clausius Equality*, CCE, as ratio $Q_H/T_H=Q_L/T_L$ for constant high- and low-temperature of the thermal reservoirs, to be the precursor for *Clausius Equality*, as circular integral for a reversible cycle with variable temperatures, $\oint \delta Q/T = 0$. Then from those correlations a new property, *entropy*, was inferred by Clausius, to be later generalized with *Clausius Inequality* as the *entropy balance*, as "quantification" of the *Second Law* of thermodynamics.

The *Carnot-Clausius Equality* (*CCE,* as finalized and renamed hear to reconcile and streamline it with the cyclic *Clausius Equality*, was named as *Carnot [Ratio] Equality*, *CtEq*, in [11]), is in essence, the entropy balance, i.e., "*entropy-in* equal to *entropy-out*" of the reversible Carnot cycle at constant in- and out-temperatures, while the *Clausius Equality* is also the balance of net-entropy (*in-minus-out*) of a reversible cycle with varying temperatures, a cyclic integral around the cycle boundary or per cycle time period. They both represent special cases of the *entropy balance* for the steady-state, stationary processes (including quasi-stationary cyclic processes), where there are no accumulation of entropy (nor any other system properties).

Note that engines are designed to run and produce power perpetually (except for necessary maintenance and repair). Therefore, their processes have to be either steady-state (stationary processes), or quasi-steady cyclic processes, often achieved by rotating or reciprocating piston-and-cylinder machinery, or any similar energy conversion devices. They both, steady-state and/or cyclic processes, do not accumulate mass and energy, but convert input to output while interacting with the energy-reservoirs, an energy source and reference sink, the latter usually a device surrounding.

**Key NOVEL-Point 6:** *The 'caloric' is transformable to work and cannot be 'extended and renamed as entropy' which is 'the final transformation'*

It is stated in *Miss Point 4* that "*Sadi Carnot could not had been thinking of 'any 'other caloric' but heat, to imply 'entropy-like quantity' as speculated by some.*" There are creative and persuasive publications but with fundamental deficiencies, that try to draw attention and establish new interpretation of *caloric* concept, first to 'cancel' its original meaning as "heat" (e.g., "*Heat is not a noun*" by Romer, *Am. J. Phys.* 2001, *69*, 107), and more recently as "*Extended Caloric Theory*" and to "*make caloric equivalent to entropy*" as well as to generalize Carnot's '*waterfall analogy*' as '*Archetype of Waterfalls*' [16].

However, some critical statements are inconsistent with thermodynamic fundamentals. Namely, "*So, 'any restoration of equilibrium in the caloric' simply means any fall of caloric through a temperature difference*" [16]" This is elusive and misplaced in general since Carnot stated that "restoration of equilibrium of caloric" by reversible extraction of motive power, as in ideal heat engine, and not "any fall of caloric" like in irreversible heat transfer from high to low temperature in heat exchangers. Sadi Carnot also stated, "*the need for reestablishing temperature equilibrium* [using adiabatic compression or expansion] *for* [isothermal/reversible] *caloric transfer* (*Key Point II*). Furthermore, *"The authors' motivation has been to make clear that an Extended Caloric Theory that allows for production of caloric, makes caloric equivalent to entropy in macroscopic thermodynamics* [16]." This is misplaced and fundamentally incorrect: For example, the *caloric*, that in part may be isentropically converted to work and reduced (as in Carnot heat engine; entropy conserved), or dissipated to lower temperature and conserved (as in heat exchanger, while entropy will not be conserved); hence the two are not the same (see also next and elsewhere).

Therefore, the *caloric*/heat is not equivalent to *entropy*, the latter is irreversible and the '*final transformation*' (it is not possible to convert entropy to anything else - not possible to 'destroy' entropy), while dissipative heat generation or conversion to/from work, is not the 'final transformation' (caloric/heat could be increased or decreased, by heat pump or heat engine, respectively, or generated by work dissipation). Furthermore, entropy is related to caloric (heat), the latter is the conjugate product of entropy (extensity) and temperature (intensity) [11]. It would be analogous and incorrect as to claim that in *waterwheel* the 'water flow work' is equivalent to 'water volume flow', since 'work' is the conjugate product of 'volume (extensity)' and 'pressure (intensity)'. Again, the caloric may be in part converted to work and vice versa (in "*thermal transformers*" i.e., heat pumps and heat engines) and thus overall increased or reduced, the latter not possible for entropy. Furthermore, *heat* (caloric) and *work* are reversibly-transformable (interchangeable as equivalent), as formalized as the "*Carnot-Clausius Heat-Work Equivalency (CCHWE)*", i.e., 'heat at high temperature' is (reversibly) equivalent, i.e., transformable, to 'work plus heat at low temperature'. As already stated, the *CCHWE* is a fundamental and autonomous physical concept, independent of any process or device [11], see also *Figure* 5.

Therefore, the *caloric* (being thermal energy) is not *entropy* (being thermal displacement/space). Certainly, conserved '*caloric*' at Carnot's time was known as present name '*heat*,' extensively used in *calorimetry* nowadays, and "*calorie*" still being used as heat unit, especially in chemistry and life.

## 5. Conclusion

The most logical and the most probable sequential developments by Sadi Carnot, regarding the *maximum efficiency of reversible cycles* and his ingenious reasoning of the *reversible contradiction impossibility,* as well as related consequences developed by his followers, and this author's novel contributions and generalizations, are presented above and summarized by the *Key Takeaways* below.

*Key Takeaways*:

1. Conservation of *caloric* misconception was probably a fortunate catalyst leading to analogy with the *waterwheel* and Carnot's hypothesis that the *motive power* of steam engine was caused and produced by the "*fall of caloric*" (reversible cooling of hot caloric) since Carnot believed that there was no "consumption" of the caloric (*Key Point I*).
2. Carnot's reasoning that "*wherever there exists a difference of temperature, motive power can be produced*" and not be wasted for 'workless' heat transfer, was the most critical and ingenious reasoning (*Key Point II*) that led to inference of the most critical concept of Carnot's discovery (*Key Point IV & V*).
3. The "*Key discovery*" ingeniously inferred by Sadi Carnot, that *Reversible cycles must all have equal and maximum efficiency*, by demonstrating that otherwise would result in creation of conserved caloric and/or perpetual motion: *"the maximum of motive power resulting from the employment of steam is also the maximum of motive power realizable by any means whatever* [1] (*Key Point V*)."
4. The selected and persistent post-misconceptions and fallacies by others are also presented as *Miss Points 1-4,* including the misconceptions that the heat transferred out of the ideal Carnot cycle at lower temperature is the "*waste heat* or *loss*" as often stated. However, it is the transformable component along with motive-power (i.e., a part of heat-work interchangeability and '*reversible equivalency*') "*useful and necessary quantity,*" required for the entropy balance and completion of a perfect, reversible cycle, see *Figures* 4-7.

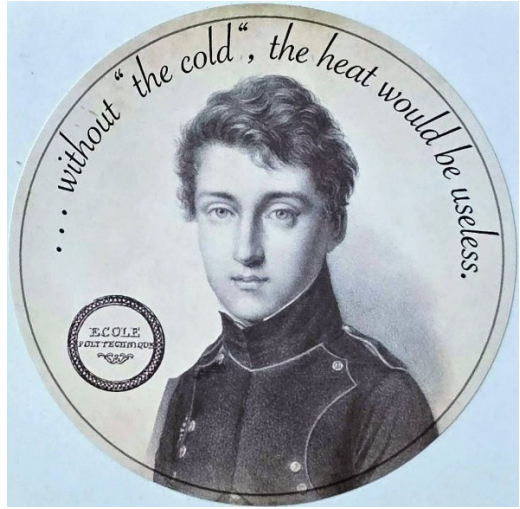

**Figure 7.** *Carnot Legacy STICKER: "...**without 'the cold', the heat would be useless**.",* École Polytechnique [3]. It illustrates claims in *Key Point I* and *Miss Point 3*: *"The [cold] heat transferred out of the Carnot cycle at lower temperature is "not a waste heat" as often stated, but it is "useful quantity", necessary for completion of the cycle." "The cold"* or heat at low temperature is actually a component, in addition to work, of the heat at high temperature. Namely, *heat* and *work* are reversibly-transformable (interchangeable as equivalent), generalized and named as *"Carnot-Clausius Heat-Work Equivalency (CCHWE)"* as *"'Heat at high temperature' is (reversibly) equivalent, i.e., transformable, to 'work plus heat at low temperature'"* The *CCHWE* is a fundamental and autonomous physical concept, independent of any process or device [11], see also *Figure* 5.

5.  The *"Reversible and Reverse"* processes and cycles are re-examined in *Key NOVEL-Point 1*. Among others, it is emphasized that the potential qualities of flux quantities (heat and different kinds of works) are not degraded but equipotentially transferred and stored between the system and its boundary surroundings, and thus conserved in every way. The spatial gradients are virtually zero at any instant while time gradients and related fluxes may be arbitrary, hence, the *time and energy rates are irrelevant for the reversible analysis* of energy balances and properties.

6.  The reversible cycle *"Maximum efficiency"* and *"Reversible equivalency"* are scrutinized in *Key NOVEL-Point 2*. The reversible efficiency implies *maximum* work extraction and *minimum* work expenditure in the *reverse* process, thereby the two must be equivalent for given conditions, thus establishing the *'Reversible equivalency'*. The reversible processes are perfect and "*equally and maximally efficient,*" and they define the concept of *"Reversible Equivalency," —* the *'true equality'* of input and output, where relevant quantities and qualities are conserved in perpetuity.

7.  *Reversible Contradiction Impossibility ("Reductio ad absurdum")* is scrutinized in *Key NOVEL-Point 3*. Namely, any reversible cycle *"under-achievement"* (getting less than maximally possible) would become an *"over-achievement"* when such cycle is reversibly 'reversed' (accomplishing with less than minimally required), and such "*'reversed' over-achievement*" would be physically impossible (would be the "*'reversible equivalency' violation*" and may violate the *conservation*

*law*s), thus establishing the *"Reversible Contradiction Impossibility"* of established fact with numerous consequences as detailed in *Key NOVEL-Point* 3 and elsewhere.

8. Sadi Carnot [1] and his followers, including Kelvin [7] and Clausius [8], ironically referred to the maximum heat-engine cycle efficiency they developed (at the time when most thermal concepts were unknown), with the *absurd conclusion*, that it does not depend on the cycle design itself nor its operation mode, therefore, the proof that *it is not the efficiency of ideal Carnot cycle 'per se'*. Therefore, their attribution is misplaced since the correlation they developed should had referred to the "maximum motive power or *'work potential'* of the thermal reservoirs" since it depends on their temperatures only, and hence, being the proof of the claim presented here, see *Key NOVEL-Point* 4 and *Figure* 6.

9. *The Clausius-Carnot [Ratio] Equality* and *Clausius Equality [Cyclic integral]* are elucidated to be the special cases of related *"Entropy [balance] boundary integral" for reversible stationary or cyclic processes,* see *Key NOVEL-Point 5*.

10. Finaly, in *Key NOVEL-Point* 6, it is reasoned why Sadi Carnot could not had been thinking of "any 'other' caloric" but heat, to imply "entropy-like quantity" as speculated by some (see also *Miss Point 4*). Furthermore, the *caloric* (being thermal energy) could not be *entropy* (being thermal displacement/space) by any stretch of the imagination as stated by some (e.g., [16]). It would be as erroneous as if water-energy in a waterwheel is claimed to be water-flow displacement. Certainly, conserved '*caloric*' at the time was known as present name '*heat',* and was used as such by Carnot.

In conclusion, even though Sadi Carnot has been often named as the "*Father of thermodynamics,"* with all farness if conceivable, it might be more appropriate for Clausius and Kelvin to be named as the *Fathers of thermodynamics*, since they meticulously developed the most critical concepts of thermodynamics, starting from thermodynamic temperature to entropy and to formulation of the *Laws* of thermodynamics, among others — whereas Sadi Carnot would be the *"Forefather of thermodynamics-to-become"* in honor of his ingenious discovery and reasoning of *heat engines reversible cycles and their maximum efficiencies* at the time when steam engines were in initial developments, when the concepts of heat and work where not fully recognized, and even the energy conservation was not established at that time.